\documentclass{moriond}

\usepackage{bm}
\usepackage{xspace}
\usepackage{amsmath}
\usepackage{tikz}
\usepackage{hyperref}

\def\be{\begin{equation}}
\def\ee{\end{equation}}
\def\bea{\begin{eqnarray}}
\def\eea{\end{eqnarray}}

\def\CP                {\ensuremath{C\!P}\xspace}

\catcode`\@=11 
\newcommand{\parenbar}{\mathpalette\p@renb@r}
\def\p@renb@r#1#2{\vbox{%
  \ifx#1\scriptscriptstyle \dimen@.7em\dimen@ii.2em\else
  \ifx#1\scriptstyle \dimen@.8em\dimen@ii.25em\else
  \dimen@1em\dimen@ii.4em\fi\fi \offinterlineskip
  \ialign{\hfill##\hfill\cr
    \vbox{\hrule width\dimen@ii}\cr
    \noalign{\vskip-.3ex}%
    \hbox to\dimen@{$\mathchar300\hfil\mathchar301$}\cr
    \noalign{\vskip-.3ex}%
    $#1#2$\cr}}}
\catcode`\@=12 

\newcommand{\mycomment}[1]{}

\newcommand{\eg}{\textit{e.\,g.}}
\newcommand{\mumu}{\ensuremath{\mu^+\mu^-}}
\newcommand{\ellell}{\ensuremath{\ell^+\ell^-}}

\newcommand{\qsq}{\ensuremath{q^2}}
\newcommand{\ctl}{\ensuremath{\cos\theta_l}}
\newcommand{\ctk}{\ensuremath{\cos\theta_K}}
\newcommand{\Bu}{\ensuremath{B^+}}
\newcommand{\Bd}{\ensuremath{B^0}}
\newcommand{\Bdb}{\ensuremath{\bar{B}^0}}
\newcommand{\Bs}{\ensuremath{B^0_s}}
\newcommand{\Lb}{\ensuremath{\Lambda_b^0}}
\newcommand{\mub}{\ensuremath{\mu{\rm b}}}
\newcommand{\mum}{\ensuremath{\mu{\rm m}}}
\newcommand{\Kp}{\ensuremath{K^+}}
\newcommand{\Km}{\ensuremath{K^-}}
\newcommand{\KS}{\ensuremath{K_{S}^0}}
\newcommand{\pip}{\ensuremath{\pi^+}}
\newcommand{\pim}{\ensuremath{\pi^-}}
\newcommand{\rhoz}{\ensuremath{\rho^0}}
\newcommand{\Kstarz}{\ensuremath{K^{*0}}}
\newcommand{\Kstarp}{\ensuremath{K^{*+}}}
\newcommand{\jpsi}{\ensuremath{J/\psi}}

\newcommand{\tev}{\ensuremath{\mathrm{\,Te\kern -0.1em V}}\xspace}
\newcommand{\gev}{\ensuremath{\mathrm{\,Ge\kern -0.1em V}}\xspace}
\newcommand{\mev}{\ensuremath{\mathrm{\,Me\kern -0.1em V}}\xspace}
\newcommand{\kev}{\ensuremath{\mathrm{\,ke\kern -0.1em V}}\xspace}
\newcommand{\ev}{\ensuremath{\mathrm{\,e\kern -0.1em V}}\xspace}
\newcommand{\gevc}{\ensuremath{{\mathrm{\,Ge\kern -0.1em V\!/}c}}\xspace}
\newcommand{\mevc}{\ensuremath{{\mathrm{\,Me\kern -0.1em V\!/}c}}\xspace}
\newcommand{\gevcc}{\ensuremath{{\mathrm{\,Ge\kern -0.1em V\!/}c^2}}\xspace}
\newcommand{\gevgevcccc}{\ensuremath{{\mathrm{\,Ge\kern -0.1em V^2\!/}c^4}}\xspace}
\newcommand{\decay}[2]{\ensuremath{#1\!\to #2}\xspace}

\newcommand{\invfb}{\ensuremath{\mbox{\,fb}^{-1}}\xspace}

\newcommand{\Photo}{\includegraphics[height=35mm]{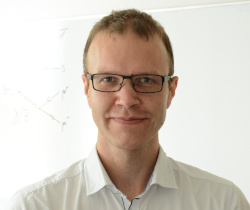}}

\begin{document}
\vspace*{4cm}
\title{Flavour Changing Neutral Current decays at LHCb}

\author{C.~Langenbruch on behalf of the LHCb collaboration}

\address{
  Physikalisches Institut,
  Heidelberg University, INF 226, 69120 Heidelberg, Germany
}

\maketitle\abstracts{
  Flavour Changing Neutral Current (FCNC) decays are forbidden at lowest perturbative order in the Standard Model (SM) and only allowed via quantum loops.
  These transitions are therefore heavily suppressed in the SM, and New Physics (NP) 
  can give significant contributions through virtual corrections.
  Of particular interest are semileptonic $\decay{b}{s(d)\ellell}$ and radiative $\decay{b}{s(d)\gamma}$ decays that allow to not only to search for the presence of NP,
  but also probe its potential operator structure through a multitude of observables.
  These observables include measurements of branching fractions, angular observables, \CP-asymmetries and tests of lepton flavour universality.
  Long-standing tensions of data with SM predictions have been observed in $\decay{b}{s\mumu}$ decays, 
  consistently in measurements of branching fractions and angular analyses. 
  However, the significance of these tensions depends on assumptions on hadronic uncertainties in the SM prediction. 
  These proceedings summarise the most recent results on FCNC decays from the LHCb experiment,
  including the legacy measurement of the key decay $\decay{\Bd}{\Kstarz\mumu}$, using $8.4\invfb$ of data from the LHC Run~1 and 2.
  }

\section{Introduction}
\begin{figure}
  \centering
  \includegraphics[height=0.19\textwidth]{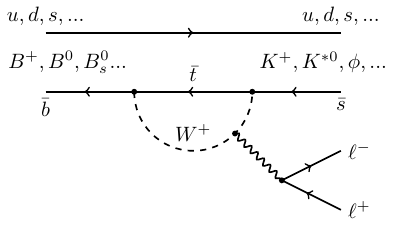}
  \includegraphics[height=0.19\textwidth]{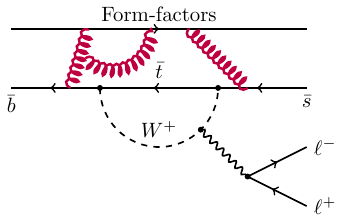}
  \includegraphics[height=0.19\textwidth]{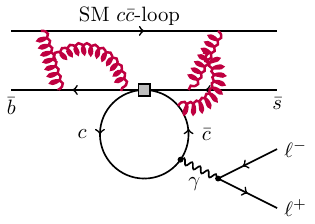}
  \caption{(Left) FCNC $\decay{b}{s\ellell}$ transitions give rise to rare $b$-hadron decays to several different final states that can be used to probe the operator structure of potential NP and check consistency between different decay modes.  The SM predictions for these processes are affected by hadronic uncertainties from (middle) local form factors and (right) non-local charm-loop contributions.\label{fig:feynmangraphs}}
\end{figure}

Rare Flavour Changing Neutral Current (FCNC) decays are sensitive probes for New Physics (NP) as they are strongly (loop-)suppressed in the Standard Model (SM).
Heavy new particles can significantly contribute to these processes via virtual corrections. 
Precision measurements of 
semileptonic $\decay{b}{s(d)\ellell}$ and radiative $\decay{b}{s(d)\gamma}$ decays therefore allow to 
perform model-agnostic searches for NP. 
Depending on the flavour-violating couplings of NP these searches can be sensitive to mass scales of ${\cal O}(100\tev)$, well beyond those accessible through direct searches.
The large $b\bar{b}$ production cross-section at the LHC of around $500\,\mub$ at $\sqrt{s}=13\tev$ results in 
enormous samples of all $b$-hadron species, including $\Bd$, $B^\pm$, and $B_s^0$-mesons, as well as $\Lb$-baryons.
As illustrated in Fig.~\ref{fig:feynmangraphs} (left) this results in a multitude of decay modes with different Lorentz structure, 
allowing both cross-checks and complementary measurements that can probe the operator structure of potential NP. 

The LHCb experiment is ideally suited for the analysis of 
rare FCNC decays 
due to its high efficiency for the reconstruction of secondary vertices from $b$-hadron decays provided by 
the LHCb Vertex Locator (VeLo), which exhibits an impact parameter resolution of around $20\,\mum$. 
The LHCb tracking system furthermore provides an excellent relative momentum resolution of $0.5\text{--}1\%$. 
In addition, LHCb has excellent particle identification capabilities due to its two Ring Imaging Cherenkov (RICH) detectors and the muon chambers.
Finally, the experiment achieves high signal efficiencies due to its low trigger thresholds,
and, since the start of Run~3, employs a flexible and highly efficient full software trigger. 

\section{Branching Fractions}
The LHCb experiment performs precision determinations of the branching fractions of rare $\decay{b}{s\mumu}$ processes
by measuring them relative to the corresponding tree-level $\decay{b}{c\bar{c}s}$ decays,
which result in the same final states via $\jpsi(\to\mumu)$ transitions. 
The SM predictions for the branching fractions of FCNC decays are unfortunately affected by significant hadronic uncertainties,
from both non-perturbative form factors, sketched in Fig.~\ref{fig:feynmangraphs} (middle), and charm-loop contributions, shown in Fig.~\ref{fig:feynmangraphs} (right). 
While there has been recent progress on form factor uncertainties \eg\ from lattice calculations~\cite{Parrott:2022zte} the contributions from the charm-loop are very challenging to estimate and the corresponding uncertainties are currently under discussion. 

\begin{figure}
  \centering 
  \includegraphics[width=0.425\textwidth]{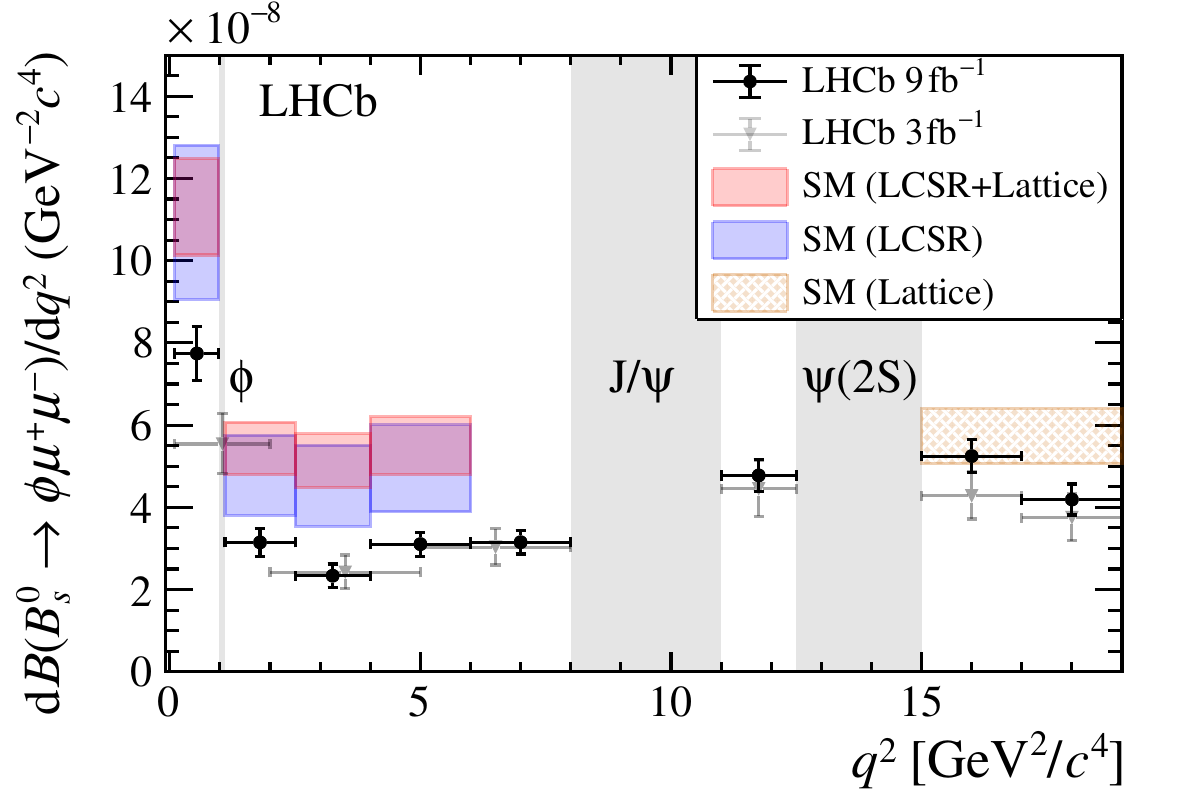}
  \hspace*{0.02\textwidth}
  \includegraphics[width=0.425\textwidth]{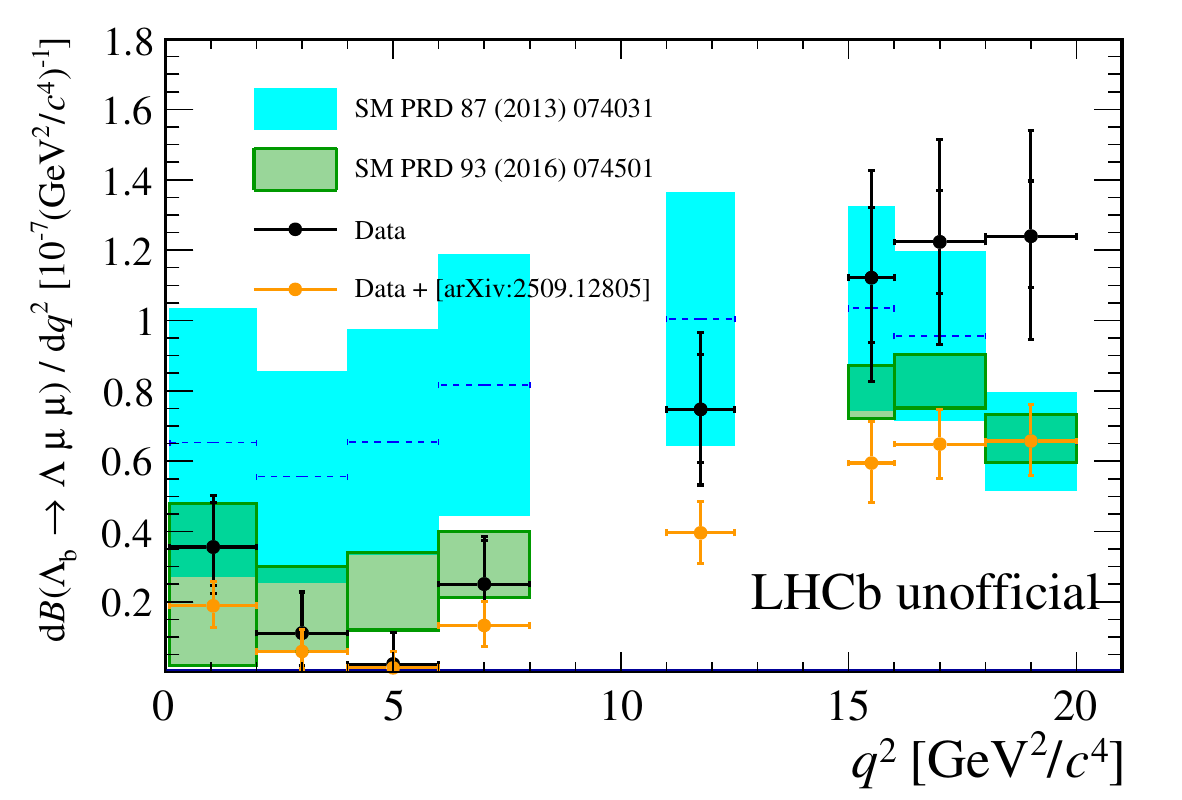}
  \caption{(Left)
    Branching fraction of the rare decay $\decay{\Bs}{\phi\mumu}$~\protect\cite{LHCb:2021zwz}.
    (Right) Branching fraction of the rare baryonic decay $\decay{\Lb}{\Lambda\mumu}$~\protect\cite{LHCb:2015tgy,LHCb:2025jva,Detmold:2012vy,Detmold:2016pkz}.
    \label{fig:branchingfractions}
}
\end{figure}

The measurements are performed in bins of $\qsq=m^2(\mumu)$, which gives some information on the relative contributions from different effective operators. 
The results for the decay $\decay{\Bs}{\phi\mumu}$ are shown in Fig.~\ref{fig:branchingfractions} (left). 
At intermediate \qsq\ values the measurement~\cite{LHCb:2021zwz} lies around $3$ standard deviations ($\sigma$)
below a SM prediction~\cite{Bharucha:2015bzk,Horgan:2015vla}.
Consistent tensions with SM predictions have also been observed in the decays $\decay{\Bu}{\Kp\mumu}$, $\decay{\Bd}{K^0\mumu}$, $\decay{\Bu}{\Kstarp\mumu}$, and $\decay{\Lb}{\Lambda^0\mumu}$~\cite{LHCb:2014cxe,LHCb:2015tgy}.
The branching fraction of the normalisation mode for $\decay{\Lb}{\Lambda^0\mumu}$, $\decay{\Lb}{\jpsi\Lambda^0}$, was recently updated by LHCb in Ref.~\cite{LHCb:2025jva}.
This lead to a significant reduction of the branching fraction of the rare mode, 
illustrated in Fig.~\ref{fig:branchingfractions} (right),
where the updated measurement is compared with two SM predictions~\cite{Detmold:2012vy,Detmold:2016pkz}. 
Low branching fractions have also been reported by the CMS experiment for $\decay{\Bd}{\Kstarz\mumu}$~\cite{CMS:2015bcy}
and 
for $\decay{\Bs}{\phi\mumu}$~\cite{CMS-PAS-BPH-23-003}.
While the tensions with the SM predictions are consistent, their significance depends on the mentioned assumptions on hadronic uncertainties. 

Most recently, the LHCb experiment has found first evidence for the $\decay{b}{s\mumu}$ decay $\decay{\Bu}{\bar{\Lambda}^0p\mumu}$, a meson decay with a baryonic final state, with $3.5\,\sigma$ significance~\cite{LHCb:2026qwd}.
The decay is studied in two regions of $m(\bar{\Lambda}^0p)$, and its branching fraction in the region $m(\bar{\Lambda}^0p)<2.8\gev$ is measured to be
${\cal B}=(1.70^{+0.65}_{-0.56}\pm 0.17 \pm 0.14)\times 10^{-8}$, 
around $2\,\sigma$ below a SM prediction~\cite{LHCb:2026qwd}. 

\begin{figure}
  \centering
  \includegraphics[width=0.425\textwidth]{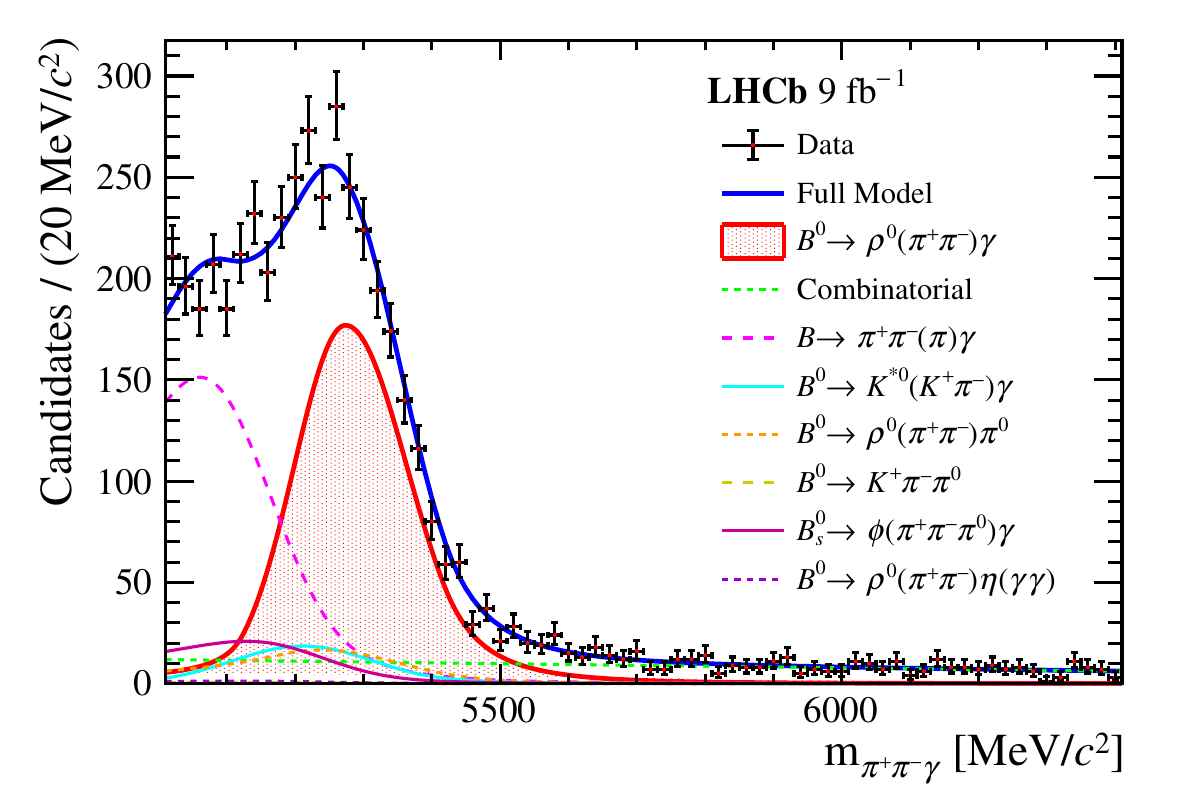}
  \hspace*{0.02\textwidth}
  \begin{tikzpicture}[inner sep=0mm, outer sep=0mm]
    \draw (0,0) node [above] {\includegraphics[width=0.425\textwidth]{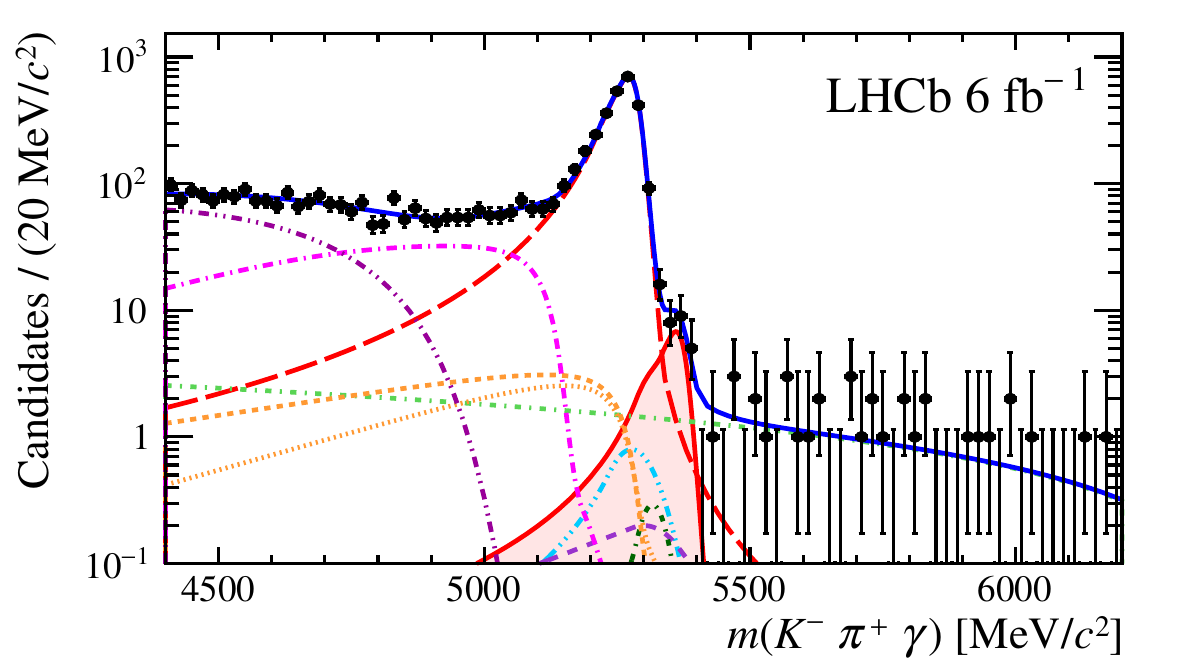}};
    \draw (0,0) node [below] {\includegraphics[width=0.425\textwidth]{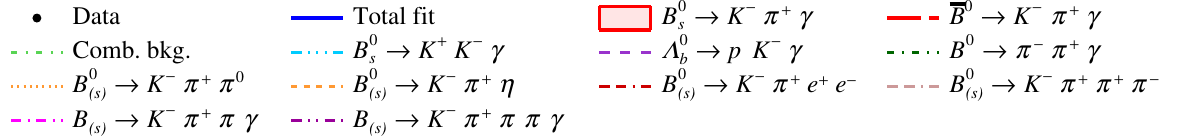}};
  \end{tikzpicture}
  \caption{The radiative $\decay{b}{d\gamma}$ decays (left) $\decay{\Bd}{\rho^0\gamma}$~\protect\cite{LHCb:2025hhv} and (right) $\decay{\Bs}{\Km\pip\gamma}$~\protect\cite{LHCb:2026udc}.\label{fig:radiative}}
\end{figure}

Furthermore, LHCb has provided measurements of two radiative $\decay{b}{d\gamma}$ decays, that, when normalising to the $\decay{b}{s\gamma}$ decay $\decay{\Bd}{\Kstarz\gamma}$,
exhibit sensitivity to the ratio of CKM matrix elements $|V_{\mathrm{td}}/V_{\mathrm{ts}}|$.
Figure~\ref{fig:radiative} (left) shows the $\decay{b}{d\gamma}$ decay $\decay{\Bd}{\rhoz\gamma}$, where the photon is reconstructed in the electromagnetic calorimeter.
The branching fraction (ratio) is determined to be ${\cal B}(\decay{\Bd}{\rhoz\gamma})=(7.9\pm 0.3\pm 0.2\pm 0.2)\times 10^{-7}$ (${\cal B}(\decay{\Bd}{\rhoz\gamma})/{\cal B}(\decay{\Bd}{\Kstarz\gamma})=0.0189\pm 0.0007\pm 0.0005$)~\cite{LHCb:2025hhv}. 
These are the world's most precise measurements and consistent with the world average. 
In addition, LHCb has found first evidence for the rare $\decay{b}{d\gamma}$ decay $\decay{\Bs}{\Km\pip\gamma}$ with a significance of $3.5\,\sigma$.
In this analysis, the photon is reconstructed using $\gamma\to e^+e^-$ conversions resulting in an improved $B$ mass resolution as shown in Fig.~\ref{fig:radiative} (right),
to allow for the separation of the $\Bs$ signal from the overwhelming $\Bd$ mass peak. 
In good agreement with the SM prediction LHCb measures ${\cal B}(\decay{\Bs}{\Km\pip\gamma})/{\cal B}(\decay{\Bdb}{\Km\pip\gamma})=(3.7\pm 1.2\pm 0.4)\times 10^{-2}$ in the
region $796<m(\Km\pip)<996\mev$~\cite{LHCb:2026udc}. 
  
\section{Angular Analyses}
Angular analyses of $\decay{b}{s\mumu}$ decays allow to precisely probe their operator structure, 
thereby providing information 
not available from decay rates. 
LHCb has recently performed a model-independent angular analysis of the flagship decay $\decay{\Bd}{\Kstarz(\to \Kp\pim)\mumu}$ in bins of $\qsq$, using $8.4\invfb$ of Run~1 and~2 data~\cite{LHCb:2025mqb}. 
The decay is fully described by five degrees of freedom, three decay angles $\vec{\Omega}=\{\ctl,\ctk,\phi\}$, as well as $\qsq$ and $m(\Kp\pim)$. 
LHCb fits the \CP-asymmetries $S_i$ and the \CP-asymmetries $A_i$ using the differential decay rate
\begin{align*}
\frac{1}{{\rm d}(\Gamma+\bar{\Gamma})/{\rm d}q^2} \frac{{\rm d}^5\parenbar{\Gamma}}{{\rm d}q^2{\rm d}m_{K\pi}{\rm d}\vec{\Omega}} = (1- \hat{F}_S)& \frac{9}{64\pi}\sum_i (S_i\pm A_i) f_i(\vec{\Omega})|{\cal BW}_P(m_{K\pi})|^2\\ 
+& \frac{1}{8\pi}\sum_j (\tilde{S}_j\pm \tilde{A}_j) f_j(\vec{\Omega}) F(m_{K\pi}),\nonumber 
\end{align*}
where the first line gives the contributions from the $\Kstarz$ P-wave and the second line contributions from the S-wave as well as P-wave/S-wave interference. 
Besides the $S_i$ basis, 
fits are also performed in the $P_i^{(\prime)}$ basis, in which form-factor uncertainties cancel to leading order~\cite{Descotes-Genon:2013vna}. 
The analysis uses more data, an improved selection, and a larger $m(\Kp\pim)$ mass range compared to previous analyses~\cite{LHCb:2020lmf}, thus significantly improving signal yields. 
Furthermore, massless leptons are no longer assumed throughout and the branching fraction of the decay is determined without model assumptions. 

\begin{figure}
  \centering
  \includegraphics[width=0.425\textwidth]{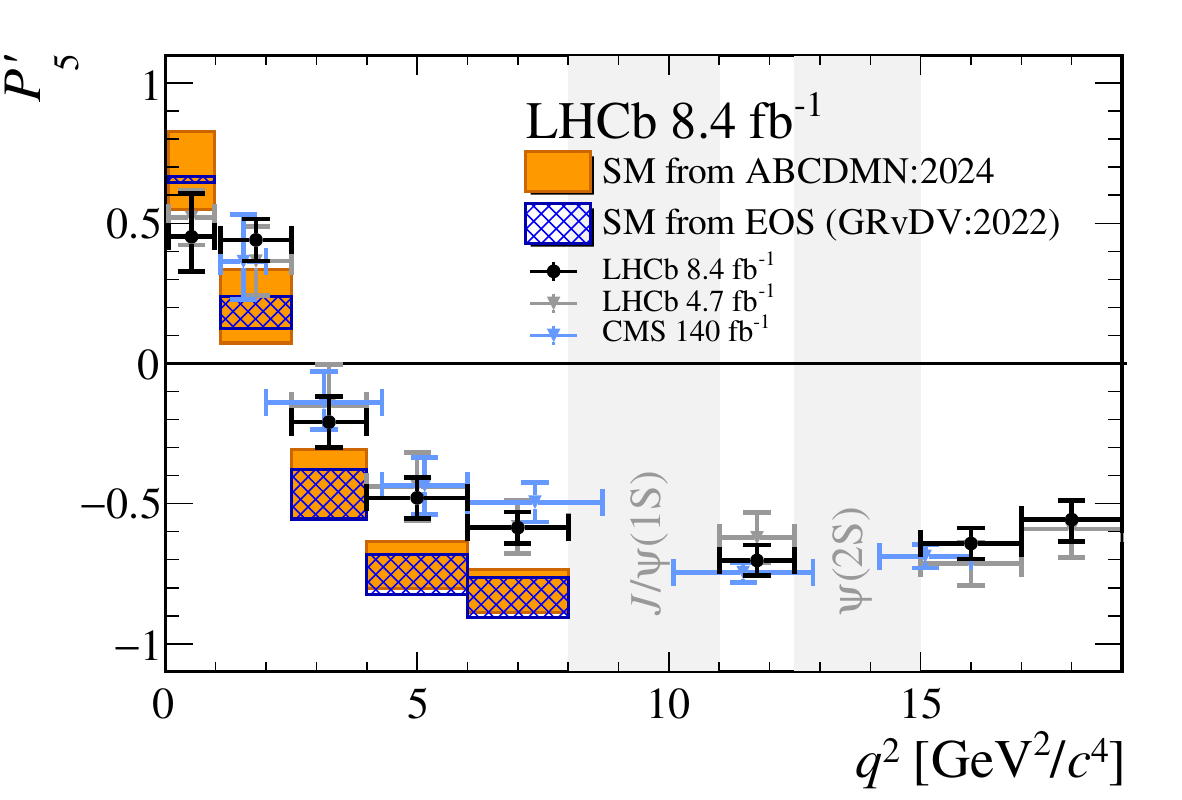}
  \hspace*{0.02\textwidth}
  \includegraphics[width=0.425\textwidth]{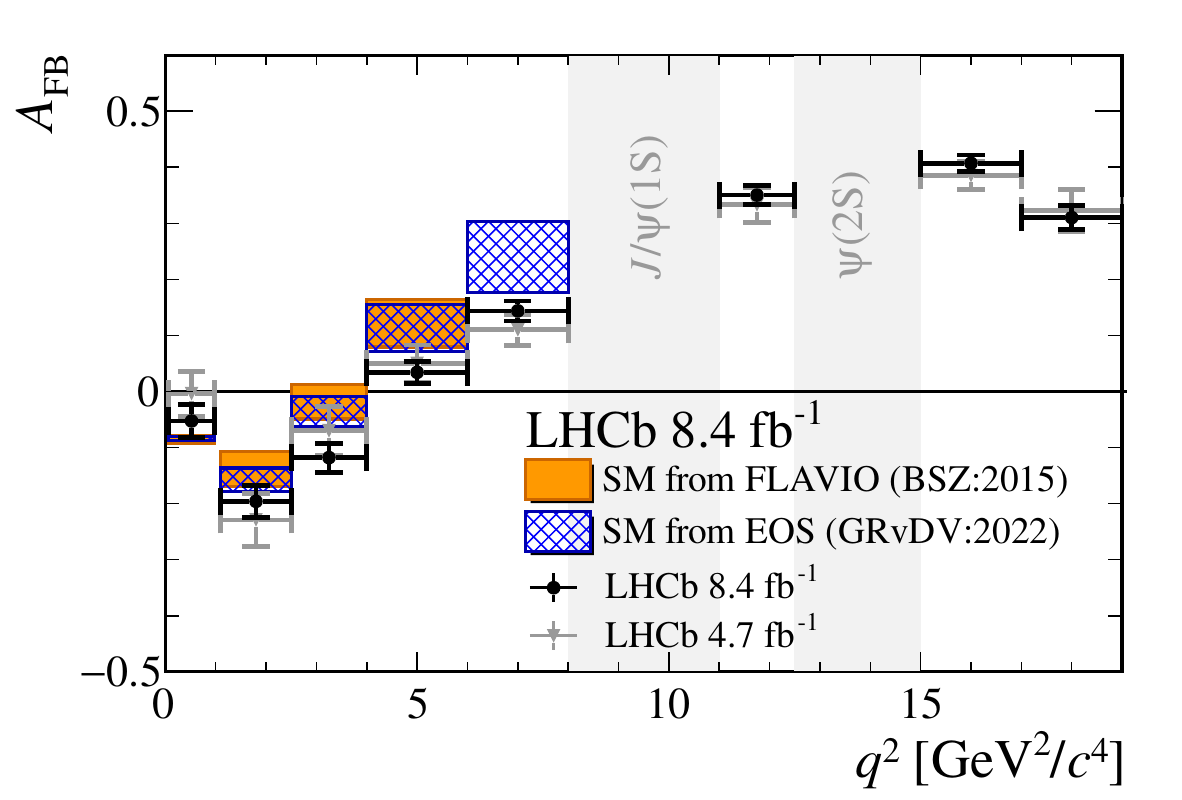}\\
  \includegraphics[width=0.425\textwidth]{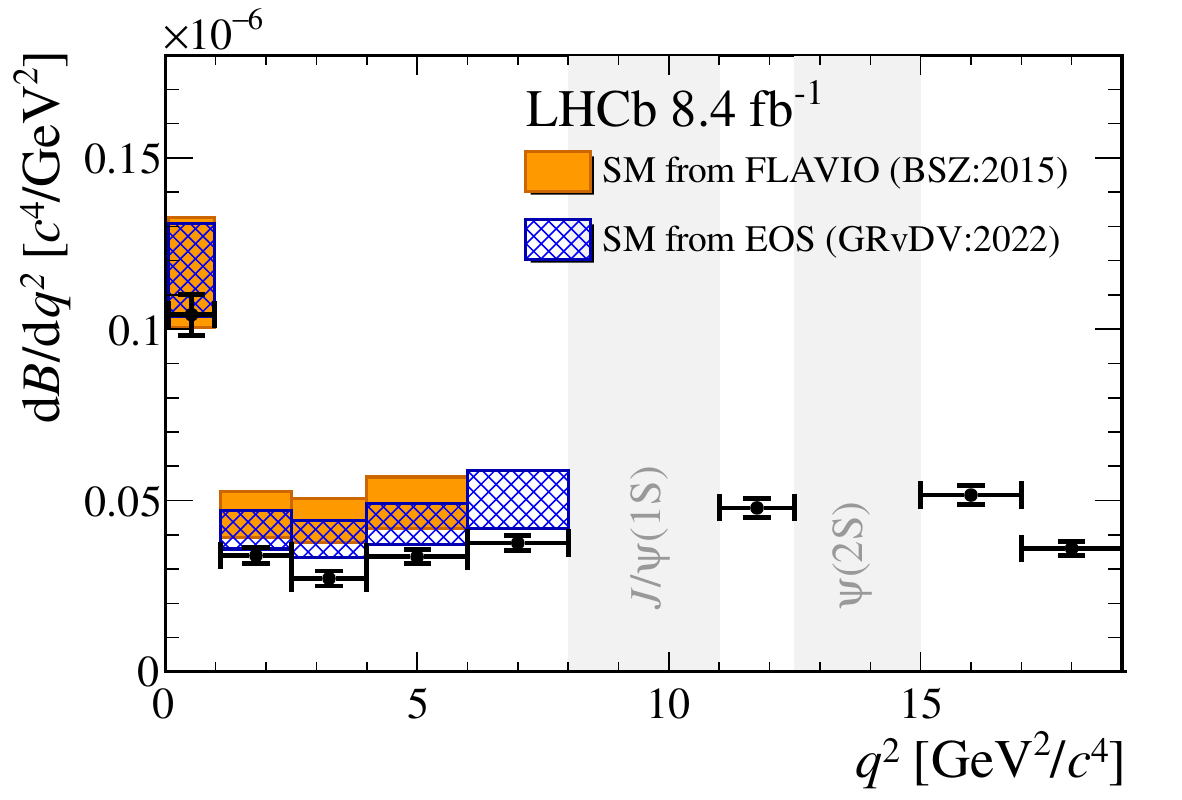}
  \hspace*{0.02\textwidth}
  \includegraphics[width=0.425\textwidth]{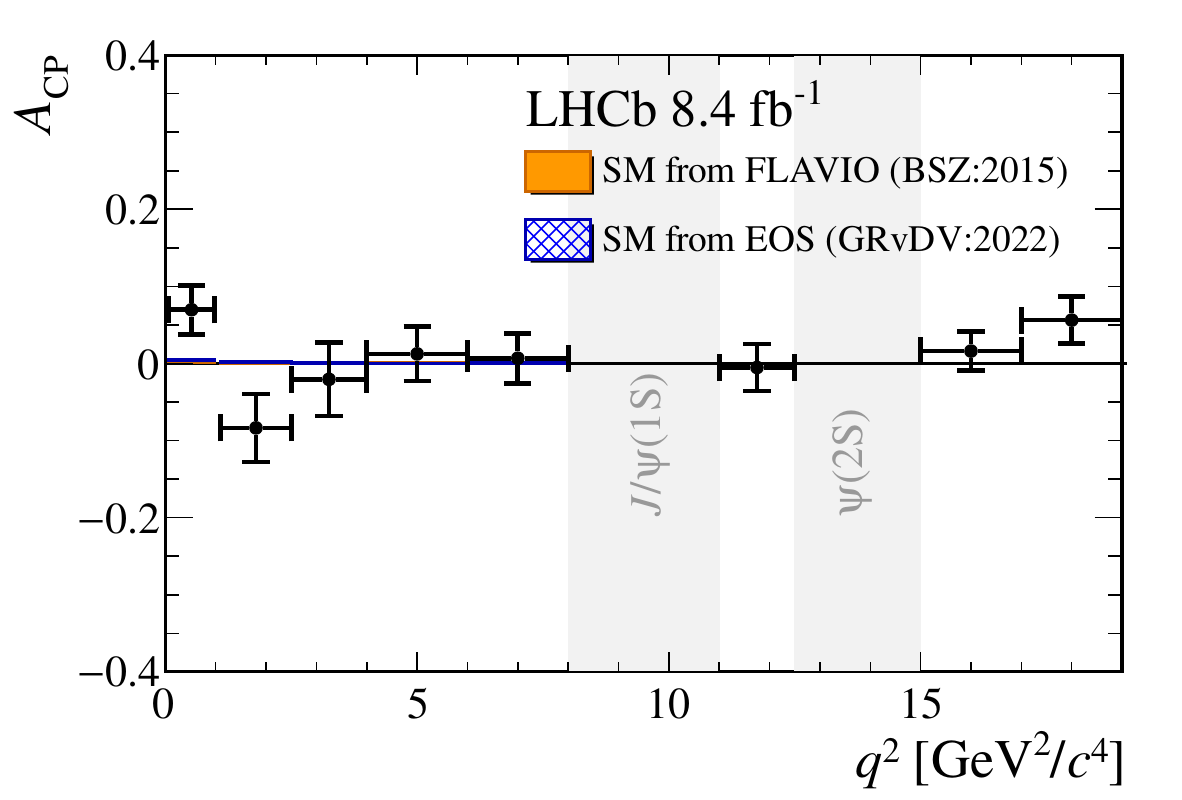}
  \caption{Results from the angular analysis of the decay $\decay{\Bd}{\Kstarz\mumu}$~\protect\cite{LHCb:2025mqb}, overlaid with SM predictions~\protect\cite{EOSAuthors:2021xpv,Gubernari:2022hxn,Alguero:2023jeh}.\label{fig:kstarmumu}}
\end{figure}

The results for the prominent angular observable $P_5^\prime$ are shown in Fig.~\ref{fig:kstarmumu} (top left).
They are in excellent agreement with previous measurements by LHCb~\cite{LHCb:2020lmf} and CMS~\cite{CMS:2024atz}, which are also shown in the figure. 
The results confirm tensions of the data with SM predictions 
at intermediate $\qsq$, corresponding to local significances of 
$2.6$ ($2.1$) and $2.7$ ($2.4$) $\sigma$ in the \qsq\ bins $[4,6]$ and $[6,8]\,\gevgevcccc$ for the prediction from Ref.~\cite{EOSAuthors:2021xpv,Gubernari:2022hxn} (Ref.~\cite{Alguero:2023jeh}).
Tensions are also visible at intermediate \qsq\ for the forward-backward asymmetry $A_{\rm FB}$, shown in Fig.~\ref{fig:kstarmumu} (top right), 
and for the branching fraction (bottom left), which lies below the SM prediction, consistent with other $\decay{b}{s\mumu}$ decays. 
Other observables are in good agreement with the SM prediction, such as the \CP-asymmetry $A_{\CP}$ shown in Fig.~\ref{fig:kstarmumu} (bottom right),
which constitutes an observable with a clean SM prediction. 

To interpret the results, a fit of the effective vector coupling ${\cal R}e C_9$ is performed, using the P-wave observables in the $\qsq$ bins below $8\gevgevcccc$ and two flavour physics software packages~\cite{Straub:2018kue,EOSAuthors:2021xpv} with different assumptions on (non-local) form factors.
The  resulting shift of ${\cal R}e C_9$ from its SM prediction is found to be $\Delta{\cal R}e C_9=-0.93^{+0.18}_{-0.16}$ ($\Delta{\cal R}e C_9=-0.94^{+0.22}_{-0.22}$)
with a significance of $4.1\,\sigma$ ($4.0\,\sigma$) using Ref.~\cite{Straub:2018kue} (Ref.~\cite{EOSAuthors:2021xpv}). 
While the angular observables are less affected by form-factor uncertainties than branching fractions,
they are however still affected by potential pollution from the charm-loop and thus no evidence for NP can be claimed despite the high significances. 

\section{\CP-Asymmetries}
\begin{figure}
  \centering
  \includegraphics[width=0.425\textwidth]{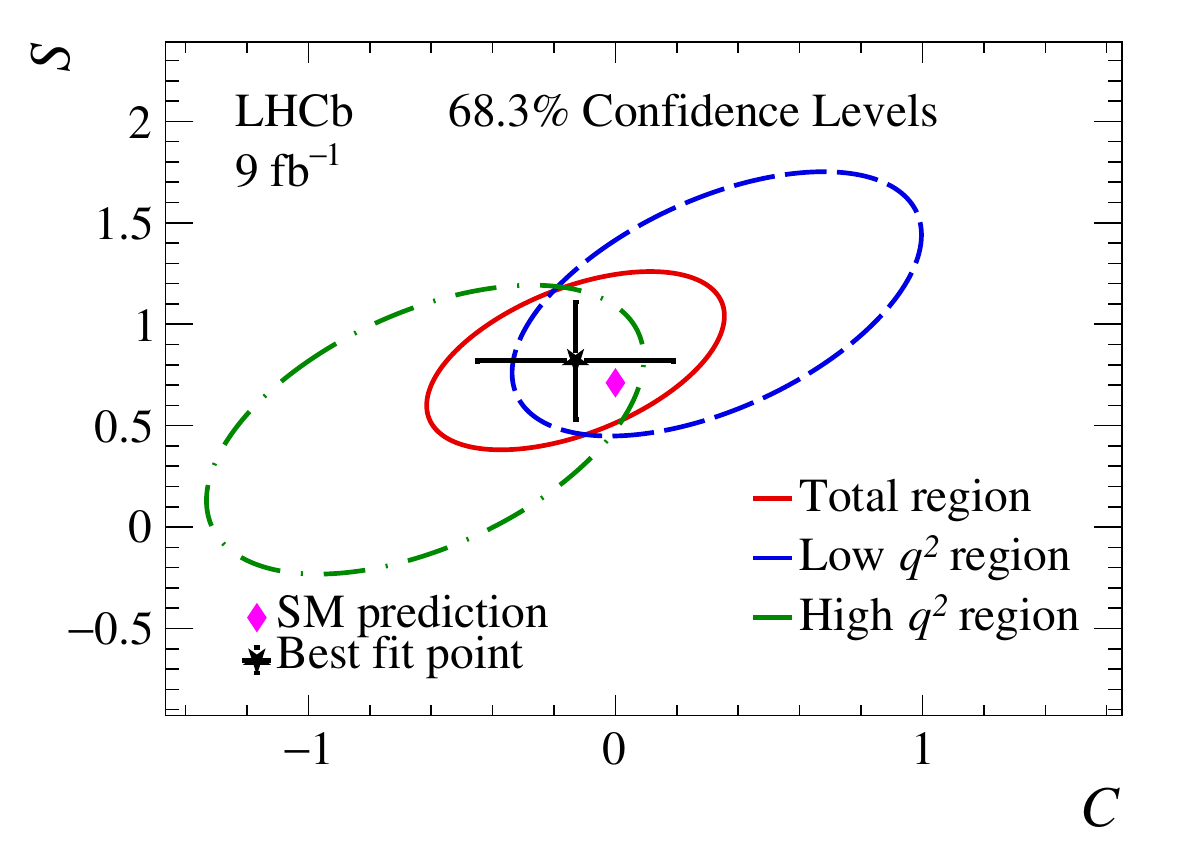}
  \hspace*{0.02\textwidth}
  \includegraphics[width=0.425\textwidth]{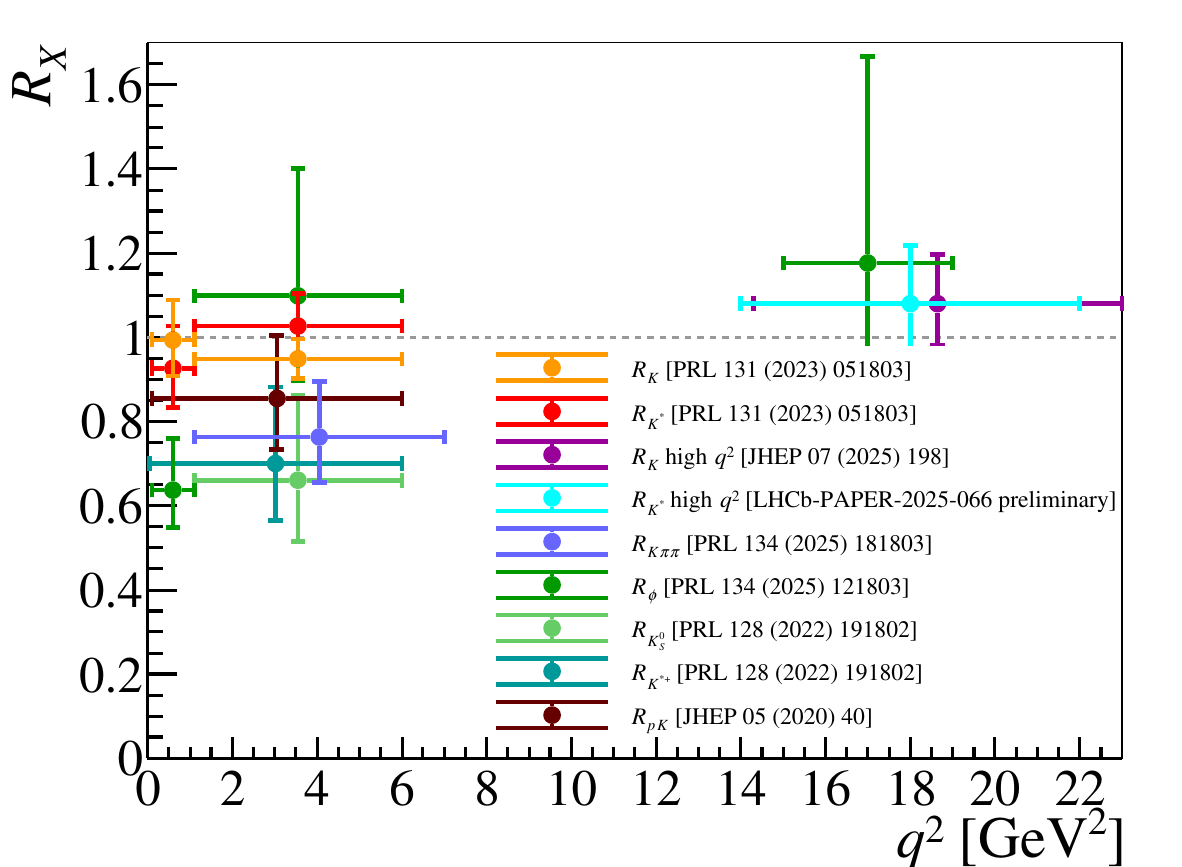}
  \caption{(Left) Clean observables include \CP-asymmetries such as those measured in the time-dependent analysis of $\decay{\Bd}{\KS\mumu}$ decays~\protect\cite{LHCb:2026ukl},
    and (right) tests of lepton flavour universality in $\decay{b}{s\ellell}$ decays.~\protect\cite{LHCb:rksthighq2,LHCb:2022qnv,LHCb:2022vje,LHCb:2025ilq,LHCb:2024yci,LHCb:2024rto,LHCb:2021lvy,LHCb:2019efc}\label{fig:clean}}
\end{figure}

\CP-asymmetries in rare $\decay{b}{s\mumu}$ decays, such as $A_{\CP}(\decay{\Bd}{\Kstarz\mumu})$ shown in Fig.~\ref{fig:kstarmumu} (bottom right) constitute clean observables,
as they are sensitive to weak phases that are not induced by QCD effects. 
LHCb has recently measured the time-dependent \CP-asymmetry ${\cal A}_{\CP}=S\sin(\Delta m_d t)-C\cos(\Delta m_d t)$ using the rare decay $\decay{\Bd}{\KS\mumu}$,
where $S$ is mixing-induced and $C$ from potential direct \CP-violation.
In the SM, the expectation is $S=\sin 2\beta_d$ and $C=0$.
Integrated over the full \qsq\ range $941\pm 55$ signal events are found and the \CP-violating parameters are determined to be $S=0.82\pm 0.29\pm 0.05$ and $C=-0.13\pm 0.32\pm 0.04$~\cite{LHCb:2026ukl},
in good agreement with the SM expectation, as shown in Fig.~\ref{fig:clean} (left). 

\section{Run~3 and Future Data Samples}
\begin{figure}
  \centering
  \includegraphics[height=0.35\textwidth]{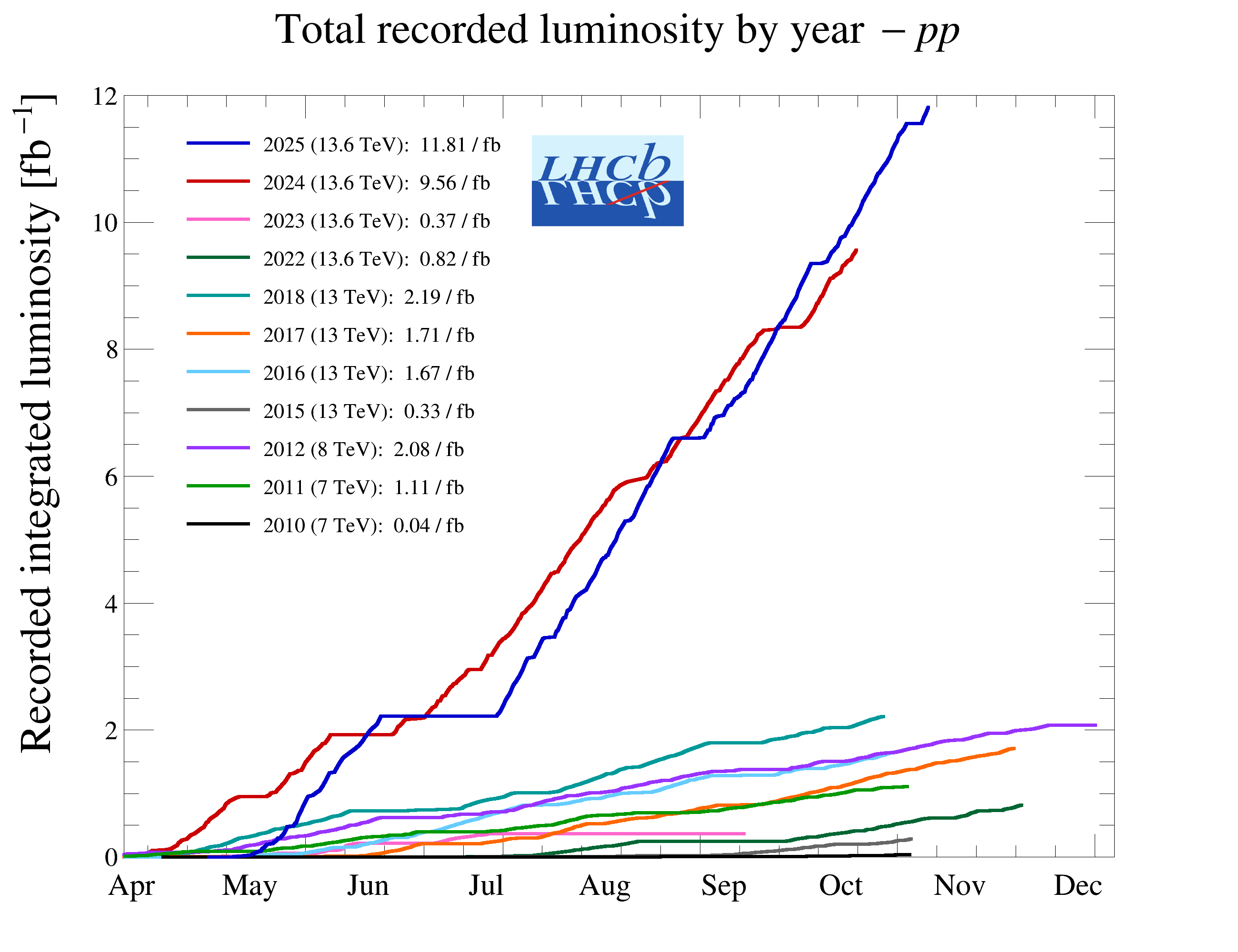}
  \hspace*{0.02\textwidth}
  \includegraphics[height=0.35\textwidth]{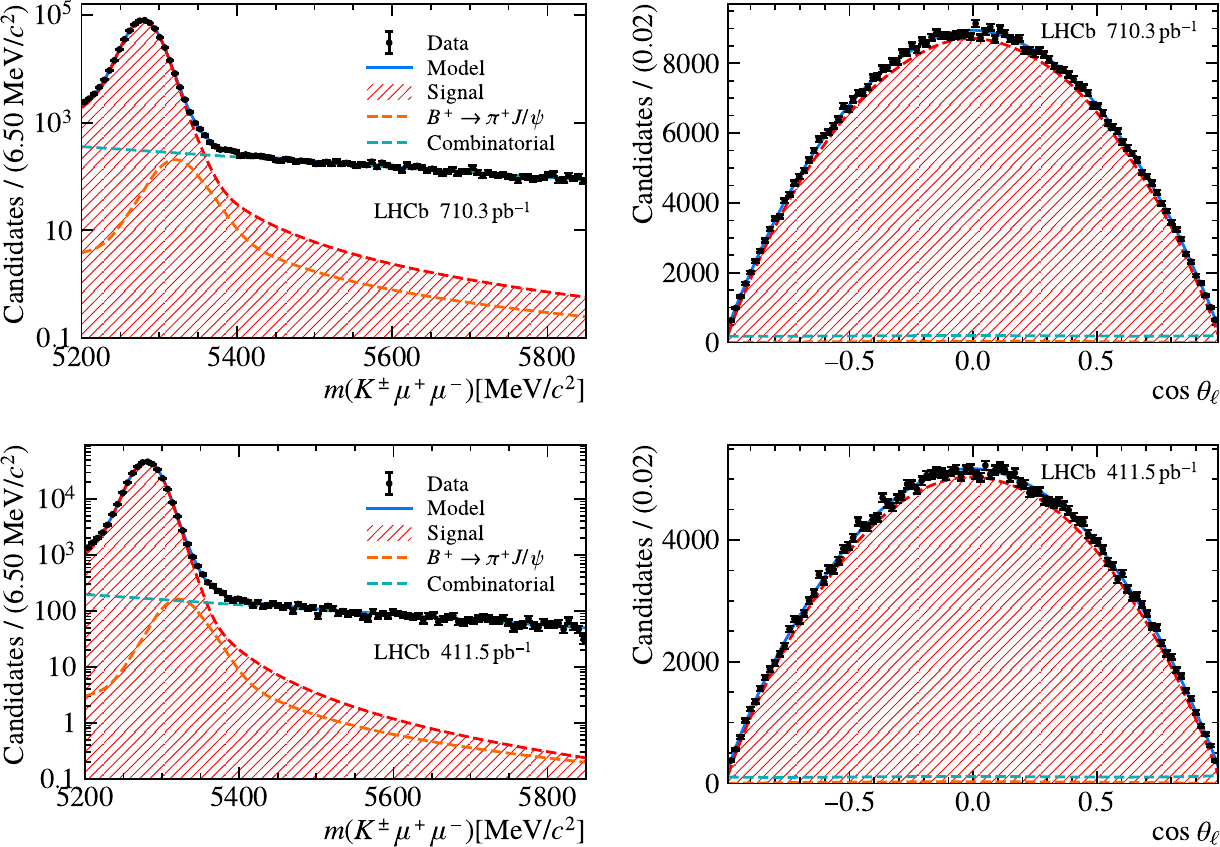}
  \caption{(Left) Data collected by the LHCb experiment in the years 2010--2025. (Right) Angular analysis of the decay $\decay{\Bu}{\Kp\jpsi(\to\mumu)}$ using $1\invfb$ of 2024 Run~3 data~\protect\cite{LHCb:2025tyz}.\label{fig:run3}}
\end{figure}

Since the start of Run~3 the LHCb experiment has collected $pp$ collision data corresponding to an integrated luminosity of more than $22\invfb$, as shown in Fig.~\ref{fig:run3} (left).
For analyses of rare decays that are statistically dominated this large data sample is extremely valuable,
but a good understanding of the data taken by the new detector is required.
To this end, LHCb has performed an angular analysis of the decay $\decay{\Bu}{\jpsi(\to\mumu)\Kp}$ using $1\invfb$ of 2024 data. 
The analysis determines the forward-backward asymmetry $A_{\rm FB}=0.00019\pm 0.00048\pm 0.00033$ and the flat parameter $F_{\rm H}=0.0005\pm 0.0011\pm 0.0014$~\cite{LHCb:2025tyz}, 
in excellent agreement with the expectation of zero.
In addition, these parameters are measured differential in bins of 17 different variables, and no significant deviations from the expectation are found, giving great confidence in the quality of the data.

\section{Lepton Flavour Universality tests}
Lepton flavour universality (LFU) can be tested in rare $\decay{b}{s\ellell}$ decays by measuring branching fraction ratios of $\decay{b}{s\mumu}$ and $\decay{b}{s e^+ e^-}$ decays. 
If lepton masses can be neglected, these ratios are unity in the SM and uncertainties from QED effects are at most of the order of $1\%$. 
Crucially, the SM prediction is not affected by hadronic uncertainties that cancel in the ratio.
Most recently, the LHCb experiment has measured $R_{K^*}$ in the high $\qsq$ region above $14\gevgevcccc$ using $9\invfb$ of Run~1 and~2 data. 
The analysis is validated using the charmonium decays $\decay{\Bd}{\jpsi(\to\ellell)\bigl(\psi(2S)(\to\ellell)\bigr)\Kstarz}$,
resulting in the control measurements $r_{\jpsi}=1.035\pm 0.013\pm 0.035$ and $R_{\psi(2S)}=1.034\pm 0.033\pm 0.012$, in excellent agreement with the expectation. 
The value of $R_{K^*}$ is found to be $R_{K^*}=1.08^{+0.14}_{-0.12}\pm 0.07$~\cite{LHCb:rksthighq2}, in good agreement with the SM prediction. 
Figure~\ref{fig:clean} (right) shows this result, together with previous measurements of LFU by the LHCb collaboration~\cite{LHCb:2022qnv,LHCb:2022vje,LHCb:2025ilq,LHCb:2024yci,LHCb:2024rto,LHCb:2021lvy,LHCb:2019efc}. 

\section{Conclusions}
\enlargethispage{2em}
These proceedings summarise recent results on FCNC decays from the LHCb collaboration. 
Most observables are in good agreement with the SM predictions, but consistent tensions persist in measurements of branching fractions and angular analyses.
These tensions are particularly prominent in the angular analysis of the key decay mode $\decay{\Bd}{\Kstarz\mumu}$,
where the tensions are found to correspond to around $4\,\sigma$ in a fit of the vector coupling ${\cal R}e C_9$. 
However, the SM predictions for branching fractions and angular observables are affected by significant hadronic uncertainties. 
In particular contributions from the so-called charm loop are 
currently under discussion. 
The clarification of these tensions will therefore require additional work, both in theory as well as in experiment. 
On the experimental side the large Run~3 data sample will allow for measurements of unprecedented precision in rare decays that are largely statistically dominated. 
First control measurements using Run~3 data illustrate the excellent quality of these data samples, that already correspond to an integrated luminosity of more than $22\invfb$. 

\section*{Acknowledgements}
C.\,L.\ gratefully acknowledges support by the 
Deutsche Forschungsgemeinschaft (DFG), 
grant identifiers LA 3937/1-1/2 and LA 3937/2-1/2. 

\section*{References}
\scriptsize
\bibliography{moriond}
\normalsize

\end{document}